\documentclass[twocolumn,prl,showpacs]{revtex4-1}
\usepackage{dcolumn}
\usepackage{amsfonts}
\usepackage{amsmath}
\usepackage{amssymb}
\usepackage{bm}
\usepackage{epsfig}
\usepackage[usenames]{color}

\begin{document}


\newcommand{\brm}[1]{\bm{{\rm #1}}}
\newcommand{\tens}[1]{\underline{\underline{#1}}}
\newcommand{\mm}{\overset{\leftrightarrow}{m}}
\newcommand{\xv}{\bm{{\rm x}}}
\newcommand{\Rv}{\bm{{\rm R}}}
\newcommand{\uv}{\bm{{\rm u}}}
\newcommand{\nv}{\bm{{\rm n}}}
\newcommand{\Nv}{\bm{{\rm N}}}
\newcommand{\ev}{\bm{{\rm e}}}
\newcommand{\Gv}{\bm{{\rm G}}}
\newcommand{\KK}{\bm{{\rm K}}}

\newcommand{\Ochange}[1]{{\color{red}{#1}}}
\newcommand{\Ocomment}[1]{{\color{PineGreen}{#1}}}
\newcommand{\Tcomment}[1]{{\color{ProcessBlue}{#1}}}
\newcommand{\Tchange}[1]{}

\title{Penrose Tilings as Jammed Solids}

\author{Olaf Stenull}
\affiliation{Department of Physics and Astronomy, University of
Pennsylvania, Philadelphia, PA 19104, USA }

\author{T. C. Lubensky}
\affiliation{Department of Physics and Astronomy, University of
Pennsylvania, Philadelphia, PA 19104, USA }

\date{\today}

\begin{abstract}
Penrose tilings form lattices, exhibiting $5$-fold symmetry and
isotropic elasticity, with inhomogeneous coordination much like
that of the force networks in jammed systems. Under periodic
boundary conditions, their average coordination is exactly
four. We study the elastic and vibrational properties of
rational approximants to these lattices as a function of
unit-cell size $N_S$ and find that they have of order
$\sqrt{N_S}$ zero modes and states of self stress and yet all
their elastic moduli vanish. In their generic form obtained by
randomizing site positions, their elastic and vibrational
properties are  similar to those of particulate systems at
jamming with a nonzero bulk modulus, vanishing shear modulus,
and a flat density of states.
\end{abstract}

\maketitle

Lattices of sites connected by central-force springs provide
useful models of mechanical systems as diverse as bridges,
elastic solids, and granular packings.  Their stability depends
critically on their average coordination number $z= 2 N_B/N_S$,
where $N_B$ is the number of bonds and $N_S$ the number of
sites in the lattice. Finite lattices with $z=z_c = 2 d -
(d(d+1))/N_S$ and lattices under periodic Boundary conditions
with $z=z_c = 2d$ are what we will call Maxwell lattices
\cite{Maxwell1864,footnote1}. They are critical networks that
lie at the boundary between being mechanically stable and
mechanically unstable, and they control the phonon structure
and elasticity of nearby stable lattices in which $z$ increases
above $z_c$ \cite{Wyartwit2005b,Wyart2005}. Maxwell Lattices
and their generalizations to include bond bending play an
important role in engineering structures
\cite{Heyman1999,Kassimali2005}, rigidity percolation
\cite{FengSen1984,JacobsTho1995}, the glass transition
\cite{Dyre2006,BerthierBir2011}, biopolymer gels
\cite{BroederszMac2014}, and randomly packed spheres near the
jamming transition \cite{LiuNag2010a,LiuNag2010b}. The latter
systems are macroscopically isotropic and are characterized by
a nonzero bulk modulus and a vanishing shear modulus, which
increases with $z>z_c$.  Small unit-cell periodic Maxwell
lattices have also been studied
\cite{SouslovLub2009,MaoLub2010,SunLub2012,KaneLub2014} but,
though like jamming systems they are characterized by lengths
and inverse frequencies that diverge as $z\rightarrow z_c^+$,
none exhibits a nonzero bulk and vanishing shear modulus at
$z=z_c$.

\begin{figure}[ptb]
\centering{\includegraphics[width=8.7cm]{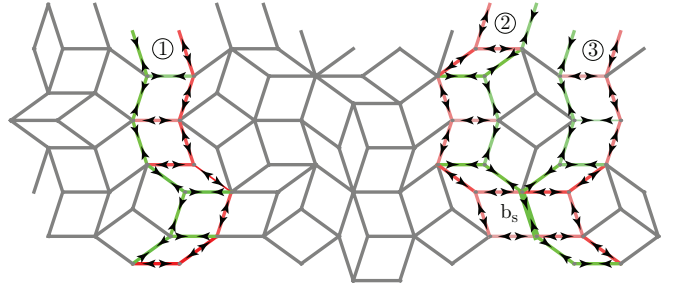}}
\caption{(color online) Unit cell of the ($1/2$)-periodic approximant of the Penrose tiling \
showing states of self stress (circled $1$, $2$, and $3$).  In all states, stress is localized
on vertical ladders with different signs of stress on opposite sides.  States $2$ and $3$ share
the bond marked $\text{b}_s$, and are not orthogonal.  They can be orthogonalized to produce states mostly, but not completely localized on the two ladders.}
\label{fig:tiling}%
\end{figure}

Here we introduce the $5$-fold symmetric quasi-crystalline
\cite{LevineSte1984,SteinhardtOst1987,DiVincenzoSte1991}
Penrose tiling \cite{Penrose74}, constructed from rhombic tiles
as shown in Fig.~\ref{fig:tiling} with local coordination
number ranging from three to as high as ten with an average of
four, as an elastically isotropic Maxwell lattice, which can be
approached via a series of rational approximants with
increasing $N_S$, each of which is a Maxwell lattice. We study
the elastic and vibrational properties \cite{QuilichiniJan1997}
of these rational approximants as the Penrose limit is
approached. All elastic moduli in all of these approximants,
like those at the rigidity percolation threshold
\cite{FengSen1984,JacobsTho1995}, are zero, and their phonon
spectra, like those of the square and kagome lattices, have of
order $\sqrt{N_S}$ zero modes. When site positions are
randomized (rendering the lattice ``generic"
\cite{GuyonCra1990,JacobsTho1996}), the properties of these
approximants are almost identical to those of jammed packings
of increasing size \cite{GoodrichNag2012}: their bulk modulus
is nonzero for all $N_S$, their shear moduli approach zero as
$N_S \rightarrow \infty$, their integrated density of states is
linear in frequency, and they exhibit quasi-localized modes
\cite{OhernNag2003}. Thus, generic Penrose tilings provide a
model system, whose construction is unhampered by dependence on
numerical protocols \cite{Dagois-Heck2012} or need for precise
equilibration \cite{GoodrichNag2012,Dagois-Heck2012}, that are
in some sense in the same ``universality class" as jammed
systems even though their detailed site-bond topology is quite
different from that of jammed systems and their method of
preparation does not guarantee a nonzero bulk modulus. We will
argue that the shear modulus must vanish in \emph{any}
isotropic Maxwell lattice with $B>0$.

We create our Penrose networks using the standard projection
procedure~\cite{deBruijn1981} from the five-dimensional
hypercube $\mathbb{Z}^5$ onto the 2-dimensional physical space.
Proper choice of the orientation of physical space within
hyperspace leads to the truly quasiperiodic rhombus tiling.
This orientation can be expressed in terms of the golden ratio
$\tau$. Approximating $\tau$ by ratios $\tau_n$ of successive
Fibonacci numbers, $\tau_1 = 1/1$, $\tau_2 = 2/1$, $\tau_3 =
3/2$, and so on, produces periodic approximants, composed of
$4$-sided rhombohedral tiles arranged in rectangular unit cells
of increasing size, that approach the $5$-fold Penrose tiling
as $n\to \infty$. The Euler relation $N_S-N_B+N_F=0$ for a
torus, where $N_F$ is the number of faces (or plaquettes),
ensures that $z=4$ under periodic boundary conditions for any
tiling with four-sided tiles. This follows because each bond is
shared by two plaquettes, implying $N_B = 2 N_F$ and $N_S = N_F
= N_B/2$. The number of sites, $N_S^n$, and bonds, $N_B^n$, in
the unit cell of the $n$th periodic approximant obey the
recursion relation
\begin{align}
N_{S, B}^{n+1} = 3\, N_{S, B}^{n} - N_{S, B}^{n-1}
\end{align}
with $N_S^1 = N_B^1 /2= 30$ and $N_S^2 = N_B^2 /2= 80$.

Periodic approximants  (Fig.~\ref{fig:tiling} shows the unit
cell of the ($1/2$)-approximant as an example) allow the
natural application of periodic boundary conditions. We study
the first eight periodic approximants ranging from $\tau_1$ to
$\tau_{8} = 34/21$, which have $30$ to $25840$ sites. Assuming
that the bonds are harmonic central-force springs, the total
elastic energy of the so-obtained Penrose network with unit
spring constant is
\begin{align}
E = \frac{1}{2} \sum_b \brm{s}_b^2 \, , \qquad \brm{s}_b = C_{bi} \brm{u}_i \, ,
\end{align}
where the sum runs over all bonds, $\brm{s}_b$ is the stretch
of bond $b$, $\brm{u}_i $ the elastic displacement of site $i$,
$C_{bi}$ the corresponding component of the compatibility
matrix $\brm{C}$ \cite{Calladine1978}.

\begin{figure}[ptb]
\centering{\includegraphics[width=6.0cm]{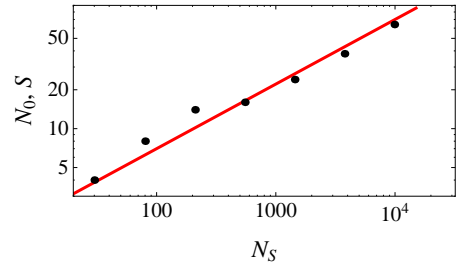}}
\caption{(color online) Number of zero modes and states of self-stress.
The (red solid) line $0.7 N_S^{1/2}$ is a guide to the eye.}
\label{fig:numZeroModes}%
\end{figure}

Along with the compatibility matrix, the equilibrium matrix
$\brm{Q} = \brm{C}^T$ relating bond tensions $t_b$ to site
forces $\brm{f}_i$ via $Q_{ib} t_b = f_i$ plays an important
role in determining the nature of modes, particularly zero
modes, in a lattice. The full vibrational spectrum is
determined by the dynamical matrix $\brm{D}= \brm{Q} \brm{C}$
(for site mass $m=1$). A zero mode is one that changes the
positions of sites but not the length of bonds, and the total
number $N_0$ of zero modes is simply the dimension of the
nullspace of $\brm{C}$. A set of bond tensions that does not
impose forces on sites is called a state of self stress
\cite{Calladine1978}, and the dimension of the null space of
$\brm{Q}$ is the number $S$ of state of self stress. The
rank-nullity theorem of linear algebra implies the generalized
Maxwell relation \cite{Calladine1978}
\begin{align}
\label{genMaxwellCount}
N_0 = 2 N_S - N_B + S \, .
\end{align}
Because $N_B = 2\, N_S$, the $5$-fold Penrose lattice and each
of its rational approximants has $N_0 = S$. This general
relation is born out by our actual calculations of eigenmodes
and states of self stress. Our results for $N_0$ and $S$ as
functions of $N_S$ are compiled in Fig.~\ref{fig:numZeroModes}.
We find that $N_0$ and $S$ roughly scale as $\sqrt{N_S}$, as is
the case in small-unit-cell periodic lattices such as the
square and kagome lattices \cite{SouslovLub2009} whose states
of self-stress arise from straight lines of bonds.

With modern computers, it is fairly straightforward to find the
set of $S$ vectors that provide a basis for the states of self
stress.  Finding a particular orthonormal basis whose elements
provide a simple geometric visualization of the origins of the
self stress is less so.  In a careful study of the $8$ states
of self stress for the $2/1$-approximant, we were able to show
that six of the states are ladder configurations, three of
which are shown in Fig.~\ref{fig:tiling}, and two were more
uniformly spread throughout the cell. Unlike the states of self
stress in the square and kagome lattice, whose tensions are all
of the same sign, these exhibit both positive and negative
tension, an important property as we shall soon see.

\begin{figure}[ptb]
\centering{\includegraphics{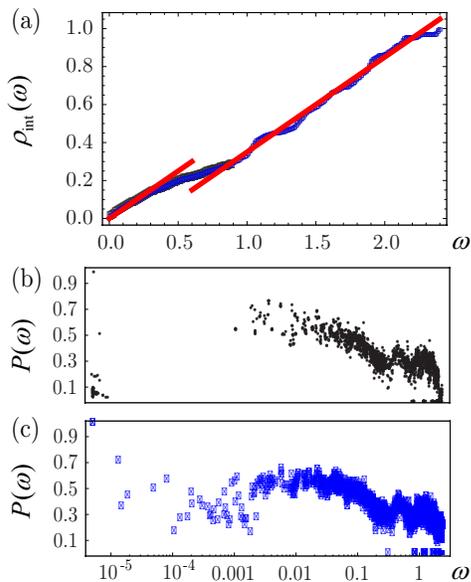}}
\caption{(color online) (a) Integrated vibrational density of states for both original and
generic lattice (indistinguishable),
(b) participation ratio of the original (black circles), and (c) the generic (blue squares)
Penrose network.  The (red solid) lines in (a) with slope $1/2$ are guides to the eye.
The zero modes in (b) have been given a small nonzero frequency for displaying
purposes so that they appear at the left boundary of the log-linear plot.}
\label{fig:intDensStatesAndPartRatio}%
\end{figure}

To study the mode frequencies and their spatial distribution,
we diagonalize the dynamical matrix $\brm{D}$ to obtain the
frequency $\omega_m$ and normalized polarization vector
$\brm{p}_{mi}$ of each site $i$ in mode $m$. From these
quantities, we extract the \emph{integrated} density of
vibrational states $\rho_\text{int} (\omega)$ and the
participation ratio
\begin{align}
P \left(  \omega_m \right) =
\frac{\big( \sum_i \left| \brm{p}_{mi} \right|^2 \big)^2}{N_S \sum_i \left| \brm{p}_{mi} \right|^4}
\end{align}
which ranges from $0$ when a mode is fully localized to $1$
when a mode is fully extended.
Fig.~\ref{fig:intDensStatesAndPartRatio} shows our results for
the $(13/8)$-approximant. The results for the other
approximants are similar. Rather than exhibiting Debye behavior
($\rho_\text{int}\sim \omega^2$) at small $\omega$,
$\rho_\text{int} (\omega) \sim \omega$, indicating a flat
non-integrated density of states as has been observed at the
jamming transition \cite{SilbertNag2005,Wyartwit2005b} as well
as in the square and kagome lattices \cite{SouslovLub2009}.
$\rho_\text{int} (\omega)$ remains approximately linear in
$\omega$ over the entire frequency range with deviations
resembling a devil's staircase \cite{Bak1982}. The
participation ratio,
Fig.~\ref{fig:intDensStatesAndPartRatio}(b) shows that the zero
modes, except for a few that include the trivial modes (rigid
translations of the network) are localized or quasi localized.
At intermediate frequencies, the modes are fairly extended, and
at higher frequencies they become more localized, again as is
the case near the jamming transition \cite{SilbertNag2005}.

States of self stress determine the linearized macroscopic
elastic energy in terms of the symmetrized strain $u_{ij}$
\cite{Pellegrino1993}.  Let the $\hat{t}_b^{\alpha}$, $\alpha =
1, ..., S$ be components of the orthnormalized basis vectors of
the nullspace of $\brm{Q}$.  Then the Elastic energy is
\begin{align}
E=\frac{1}{2} \sum_{\alpha} (\sum_b t_b^{\alpha} s_b^a )^2
\equiv \frac{1}{2} K_{ijkl} u_{ij} u_{kl},
\label{efectiveEnSS}
\end{align}
where $K_{ijkl}$ is the tensor of elastic moduli (times the
cell volume) and $s_b^a =\hat{e}_{bi} u_{ij} \hat{e}_{bj}$ is
the affine stretch of bond $b$ where $\hat{e}_b$  is the unit
vector along that bond. Since there are three independent
strains, $K_{ijkl}$ can be expresses as a $3\times 3$ elastic
matrix $\KK$ (in the Voight notation) with six independent
components leading in general to six independent elastic
constants (in two dimensions). For complete elastic stability,
all three eigenvalues of the Voight elastic matrix must be
positive. From Eq.~(\ref{efectiveEnSS}), it is clear that there
must be at least three ($d(d+1)/2$ in $d$-dimensions) states of
self stress for complete elastic stability. With $S \propto
\sqrt{N_S}$, there are more than enough states of self stress
to completely stabilize elastic distortions of the non-generic
Penrose approximants. Nevertheless, all elastic moduli are zero
for each of them. This is because the overlap of each of their
states of self-stress, which have both positive and negative
tensions [Fig.~\ref{fig:tiling}], with the affine bond strains
is zero. This is in contrast to the kagome lattice, for
example, which has three states of self stress at zero wavenum
ber and an elastic matrix with three positive eigenvalues
\cite{SunLub2012}.

Particulate systems at the jamming transition are amorphous and
generic and hence, in general, free of any but the $d$
translational zero modes (and ``rattlers", which can be
removed). To model this property in Penrose tilings, we follow
Ref.~\cite{JacobsTho1995,JacobsTho1995} and introduce small
random site displacements to produce ``generic networks" with
random bond lengths and bond angles without affecting local
topology. These local distortions eliminate all but the two
translational zero modes and their two associated states of
self-stress and thus make the generic Penrose network a more
realistic model for amorphous materials such as jammed matter.

To study the connection of the Penrose network with the jamming
transition in greater depth, we consider in the following not
only generic Penrose networks with $N_B = 2 N_S$, but also
versions of them with 1 or 2 bonds removed or 1 added somewhere
in the network. $\Delta = N_B - 2 N_S$ measures the under or
over coordination relative to the Maxwell lattice. From the
eigenmodes and states of self stress, we find that $N_0 =2$
(the trivial modes) independent of $N_S$ and $\Delta$, and $S =
2 + \Delta$ in full agreement with relation
(\ref{genMaxwellCount}). The integrated density of states for
the generic network with $\Delta = 0$ is practically
indistinguishable from that of the non-generic case, see
Fig.~\ref{fig:intDensStatesAndPartRatio} (a). In particular, we
once again obtain jamming-like $\rho_\text{int} (\omega) \sim
\omega$ for small $\omega$. The participation ratio for
$\Delta=0$, of course, differs from that of the non-generic
case, because there are fewer zero modes. In
Fig.~\ref{fig:intDensStatesAndPartRatio} (b), we see that all
but the trivial zero modes get lifted to finite-frequency,
quasi-localized  modes. In the vicinity of the jamming
transition, the non-trivial low-frequency modes are also
quasi-localized albeit the number of quasi-localized modes
relative to the total number of modes is significantly higher
there~\cite{XuNag2010}. \Tchange{Do we need to say this?}

\begin{figure}[ptb]
\centering{\includegraphics{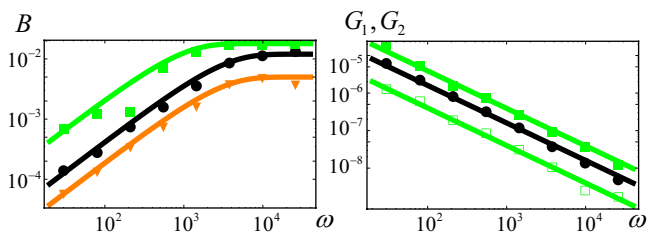}}
\caption{(color online) (a) Bulk and (b) shear moduli of the
generic Penrose network for $\Delta = 1$ (green squares),
$\Delta = 0$ (black circles), and  $\Delta = - 1$ (orange triangles). In (b),
the solid and open symbols pertain to $G_1$ and $G_2$, respectively.
The curves in (a) are fits of the form $B = \rho [1 - \exp (- \tau N_S)]$,
and the lines in (b) are fits of the form $G_1 = \chi N_S^{-1}$ and
likewise for $G_2$ with fit-parameters $\rho$, $\tau$, $\chi$ depending on $\Delta$.}
\label{fig:averagedModuli}%
\end{figure}

Next, we calculate the bulk and shear moduli of the generic
Penrose network.  Because we expect the average generic network
to exhibit isotropic elasticity in the $N_S \rightarrow \infty$
limit, we express strain in terms of its compression,
pure-shear, and simple-shear components $U=(u_{xx}+u_{yy},
u_{xx} - u_{yy}, 2 u_{xy})$, and we calculate the $3\times 3$
elastic matrix $\KK$ in this basis for each random
configuration. In the isotropic limit, this matrix becomes
${\rm diag}(B,G,G)$ where $B$ is the bulk modulus, and $G$ is
the isotropic shear modulus.  We verify that all of the states
of self stress in the network have non-zero overlap with the
vector of affine strains for $\Delta = -1,0,1$.  We then
calculate the eigenvalues of $\KK$ for each random
configuration and verify that $2+\Delta$ of them are positive.
The largest eigenvalue in all cases very quickly corresponds
with increasing $N_S$ almost exactly to pure compressions, and
we identify this eigenvalue with the bulk modulus.  When
$\Delta = -2$, all moduli are zero, when $\Delta = -1$, the
bulk modulus is the only nonzero eigenvalue. When $\Delta = 0$,
there is a second positive eigenvalue, $G_1$ which corresponds
to some combination of pure and simple shear depending of the
random configuration; and when $\Delta = 1$, there are two
shear moduli $G_1$ and $G_2$.  In the $N_S\rightarrow \infty$,
isotropy requires $G_1 = G_2$.  Thus, when $\Delta = 0$, $G_1$
must tend to zero with $N_S$ because $G_2$ is identically zero. This is a general property of periodic Maxwell lattices that approach
isotropy with increasing unit-cell-size.
Adding an extra bond does generally does not cause a discontinuous changes in
this picture, and if it does not, both $G_1$ and $G_2$ approach zero with
$N_S$.
Figure \ref{fig:averagedModuli} displays the averages of
$B$, $G_1$, and $G_2$ over configurations as a function of
$N_S$ for small for $\Delta = -1,0,1$.  At small $N_S$, $B$
undergoes a changes of about a factor of $10$ from $\Delta =
-1$ to $\Delta = 1$. As $N_s$ increases, $B$ increases in all
cases, reaching a plateaus at large $N_S$ with $B(\Delta =1)
\approx B(\Delta = 0)$ and $B(\Delta = -1)$ a factor of about
$4$ smaller. In all cases, the shear modulus approaches zero as
$1/N_S$. For completeness, we note that we also looked at
generic Penrose networks with $\Delta$ beyond $1$ and observed
elastic moduli that were qualitatively the same as for
$\Delta=1$. We also calculated the average $\langle \KK
\rangle$ of the elastic matrix over all configurations for each
$\Delta$ and $N_S$. Because of the nonlinear relation between
$\KK$ and random displacements, $\langle \KK \rangle$ generally
exhibited $3$ rather than the $2 + \Delta$ positive eigenvalues
exhibited by each configuration of $\KK$. The difference
between the average of bulk moduli calculated from $\KK$ and the
bulk modulus of the $\langle \KK \rangle$ was not generally
significant but that for the shear moduli was.

Our counting based on the index theorem
[Eq.~(\ref{genMaxwellCount})] and the relation between the
states of self stress and the elastic energy
[Eq.~(\ref{efectiveEnSS})] agrees with those obtained in the
context of jamming through requiring these packings to be
stable with respect to shape as well as volume change
\cite{Dagois-Heck2012} and through studies of finite size
effects in these packings
\cite{GoodrichNag2012,GoodrichNag2014}.  In all cases, the bond
excess $ \Delta = N_B - 2N$ in $2d$ generic latices under
periodic boundary conditions required to ensure stability with
respect to volume change only, to volume change and one shear,
and to all uniform elastic distortions is $\Delta = -1,0,1$,
respectively.  These results are equivalent to the observation
that a lattice under periodic boundary conditions cannot be
both statically and kinematically determinate
\cite{GuestHut2003}.

Penrose tilings and their rational approximants are Maxwell
lattices that approach elastic isotropy as $N_S \rightarrow
\infty$.  Remarkably, in their original form, all of their
elastic moduli are zero, but in their randomized generic form,
they, like packed spheres near jamming,  have nonzero bulk
moduli and shear moduli that vanish as $N_S \rightarrow \infty$
in spite of their not being specifically constrained to support
isotropic loads.  Their vibration eigenmodes are also similar
to those at the jamming transition.  We have not carried out
systematic investigation of the effects of adding an extensive
number of bonds, but we expect that doing so will have an
effect similar to increasing $z$ near jamming.  Our study
focussed on the systems with single unit cells under periodic
boundary conditions, i.e.,  restricting our attention to the
zero-wavenumber limit of periodically repeated unit cells. The
latter has modes at all wavenumbers with zero modes at each
wavenumber in the nongeneric lattices. The original Penrose
tilings are critical lattices, similar in many respects to the
kagome lattice with $\sim \sqrt{N_S}$ states of self stress
that can be removed by even infinitesimal displacements of
sites. This raises the possibility that controlled
displacements could lead to different topological classes
\cite{KaneLub2014} with associated zero surface modes.
Three-dimensional Penrose tilings \cite{LevineStein1986} are
also Maxwell lattices, and they should have properties similar
to their $2d$ cousins. Finally, we note that quasicrystals lie
at the boundary between periodic crystals and glasses, and it
is intriguing that slight randomization of position of lattice
sites leads to glassy-like behavior.

We thank C. P. Goodrich  and A. J. Liu for helpful discussions.
This work was supported by the NSF under grants DMR-1104707 and
DMR-1120901.

\end{document}